\documentclass[11pt,english]{smfart}

\usepackage[T1]{fontenc}
\usepackage[english]{babel}

\usepackage{amssymb,url,smfthm}
\usepackage{tcolorbox}
\usepackage{solarized-light}
\usepackage{sourcecodepro}

\lstdefinestyle{kymatio}{
  language=python,
  basicstyle=\tiny\ttfamily,
  numbers=left,
  stepnumber=1,
  numbersep=10pt,
  tabsize=4,
  showspaces=false,
  showstringspaces=false,
}
\newcommand{\BibTeX}{{\scshape Bib}\kern-.08em\TeX}
\newcommand{\T}{\S\kern .15em\relax }
\newcommand{\AMS}{$\mathcal{A}$\kern-.1667em\lower.5ex\hbox
        {$\mathcal{M}$}\kern-.125em$\mathcal{S}$}

\title{Musical Metamerism \\ with Time--Frequency Scattering}
\author{Vincent Lostanlen and Han Han\\November 2024}

\address{Nantes Université, École Centrale Nantes, CNRS, LS2N, UMR 6004, F-44000 Nantes, France }
\email{vincent.lostanlen@cnrs.fr}
\urladdr{https://audio.ls2n.fr}

\begin{document}
\def\smfbyname{}

\begin{abstract}
The concept of metamerism originates from colorimetry, where it describes a sensation of visual similarity between two colored lights despite significant differences in spectral content.
Likewise, we propose to call ``musical metamerism'' the sensation of auditory similarity which is elicited by two music fragments which differ in terms of underlying waveforms.
In this technical report, we describe a method to generate musical metamers from any audio recording.
Our method is based on joint time--frequency scattering in Kymatio, an open-source software in Python which enables GPU computing and automatic differentiation.
The advantage of our method is that it does not require any manual preprocessing, such as transcription, beat tracking, or source separation.
We provide a mathematical description of JTFS as well as some excerpts from the Kymatio source code.
Lastly, we review the prior work on JTFS and draw connections with closely related algorithms, such as spectrotemporal receptive fields (STRF), modulation power spectra (MPS), and Gabor filterbank (GBFB).
\end{abstract}

\maketitle

\tableofcontents

\section{Introduction}

\subsection{Problem statement}
What makes music memorable?
The ability of human listeners to recognize a song ``by ear'' may be easy to observe, it remains difficult to explain \cite{honing2021unravelling}.
For example, a familiar song may be identified after having undergone a certain amount of frequency transposition or time stretching.
Likewise, prior research in cognitive science has demonstrated that songs may retain their familiarity after noise vocoding---a digital audio effect which locally alters spectral content while leaving the temporal envelope essentially unchanged \cite{li2023pitch}.

Together, these findings hint at the idea that music familiarity is not reducible to the recognition of rigid patterns in the time--frequency domain.
Instead, Henkjan Honing \emph{et al.} have suggested that listeners attend to the global ``contour'' of a musical piece when perceiving it as familiar \cite{honing2018musicality}.
Here, the notion of contour is an umbrella term for various auditory attributes: i.e., melodic, harmonic, rythmic, timbral, textural, and so forth.
The commonality between all these components of contour perception resides in their invariance to local alterations of the short-term power spectrum, such as those mentioned earlier: noise vocoding, time stretching, and frequency transposition.
Thus, Honing's hypothesis, which we refer to as ``multicomponent contour perception'', is consistent with the available evidence on music familiarity \cite{honing2015without}.

Yet, there is a gap in scholarship concerning the relative weight of contour components in the cognition of familiar music pieces.
Measuring these weights experimentally is challenging, for lack of suitable ``minimal pairs'': i.e., pairs of auditory stimuli which differ from each other just enough to significantly reduce perceived familiarity while having some contour components in common \cite{honing2006evidence}.
The prospect of synthesizing such minimal pairs as auditory stimuli raises a difficult challenge for signal processing research, but one that is worth pursuing because of its application potential.

\subsection{Contribution}
In this report, we present a new method for stimulus design in music cognition experiments.
Our method relies on a nonlinear multiresolution approximation algorithm known as joint time--frequency scattering, or JTFS for short.
JTFS decomposes the spectrogram into wavelet subbands at various temporal rates and frequential scales while retaining a fine localization of coefficients in the time--frequency domain.
The Euclidean distance between two JTFS feature vectors is differentiable with respect to the underlying digital audio samples, thus lending itself to signal reconstruction via gradient descent.

Following Feather \emph{et al.}, we refer to the outcome of this reconstruction by ``metamer'' \cite{feather2019metamers,feather2023model}; i.e, a signal that is perceptually indistinguishable from the reference even so the underlying waveforms may significantly differ.
Although the synthesis of JTFS metamers has found applications in computer music \cite{lostanlen2019shape}, the same cannot yet be said of music cognition research.

The key idea is to coarsen the spectrotemporal discretization of JTFS coefficients by means of Gaussian low-pass filtering over the time and frequency variables.
After this coarsening, Euclidean distances in JTFS feature space are invariant to small shifts up to some time scale $T$ and to small transposition up to some frequency interval $F$.
Such a property of invariance aligns with the ``contour hypothesis'' of Honing \emph{et al.}, i.e., the claim that music listeners primarily attend to relative variations in spectrotemporal energy distribution in the time--frequency domain as opposed to absolute time--frequency cues.
Therefore, resynthesizing a waveform from coarsened JTFS coefficients produces a kind of ``musical metamer'', that is, a signal which resembles the original in terms of summary statistics over spectrotemporal regions of duration of the order of $T$ and pitch register of the order of $F$.


\section{Method}
This section reproduces some elements from \cite{lostanlen2019shape}, with minor adaptations.

\subsection{Joint time--frequency scattering}

Time--frequency scattering results from the cascade of two stages: an auditory filterbank and the extraction of spectrotemporal modulations with wavelets in time and log-frequency.
First, we define Morlet wavelets of center frequency $\lambda > 0$ and quality factor $Q$ as
\begin{equation}
\boldsymbol{\psi}_{\lambda}(t) =
\lambda
\exp\left(- \dfrac{\lambda^2 t^2}{2Q^2}\right) \times
( \exp(2\pi \mathrm{i} \lambda t) - \kappa),
\label{eq:psi}
\end{equation}
where the corrective term $\kappa$ ensures that each $\boldsymbol{\psi}_\lambda (t)$ has one vanishing moment, i.e., a null average.
Within a discrete setting, acoustic frequencies $\lambda$ are typically of the form $2^{n/Q}$ where $n$ is integer, thus covering the hearing range.
For $\boldsymbol{x}(t)$ a finite-energy signal, we define the CQT of $\boldsymbol{x}$ as the matrix $\mathbf{U}_1 \boldsymbol{x}(t, \lambda) =
\left \vert \boldsymbol{x} \ast \boldsymbol{\psi}_{\lambda} \right \vert (t)$; which stacks convolutions with all wavelets $\boldsymbol{\psi}_\lambda (t)$ followed by the complex modulus nonlinearity.

Secondly, we define Morlet wavelets of respective center frequencies $\alpha > 0$ and $\beta \in \mathbb{R}$ with quality factor $Q=1$.
With a slight abuse of notation, we denote these wavelets by $\boldsymbol{\psi}_\alpha(t)$ and $\boldsymbol{\psi}_\beta(\log \lambda)$ even though they do not necessarily have the same shape as the wavelets $\boldsymbol{\psi}_{\lambda}(t)$ of Equation \ref{eq:psi}. Frequencies $\alpha$, hereafter called amplitude modulation \emph{rates}, are measured in Hertz (Hz) and discretized as $2^n$ with integer $n$.
Frequencies $\beta$, hereafter called frequency modulation \emph{scales}, are measured in cycles per octave (c/o) and discretized as $\pm 2^n$ with integer $n$.
The edge case $\alpha = 0$ corresponds to $\boldsymbol{\psi}_{\alpha}(t)$ being a Gaussian low-pass filter $\boldsymbol{\phi}_T (t)$ of temporal width $T^{-1}$, measured in Hertz.
Likewise, the edge case $\beta = 0$ corresponds to $\boldsymbol{\psi}_{\beta}(\log \lambda)$ being a Gaussian low-pass filter $\boldsymbol{\phi}_F (\log \lambda)$ of bandwidth $F^{-1}$, measured in cycles per octave.
These modulation scales $\beta$ play the same role as the \emph{quefrencies} in a mel-frequency cepstrum.

\begin{figure}
\centerline{\includegraphics[width=0.75\linewidth]{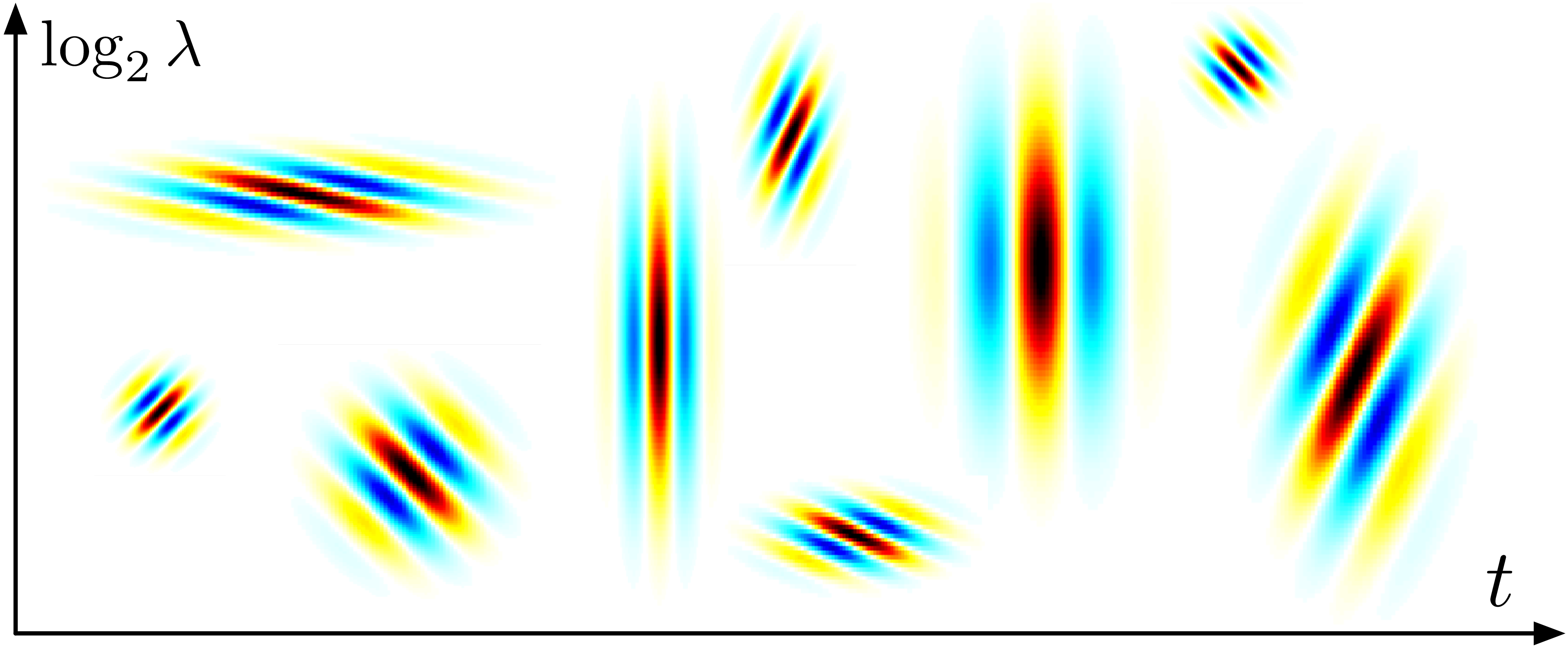}}
\caption{Interference pattern between wavelets $\boldsymbol{\psi}_\alpha (t)$ and $\boldsymbol{\psi}_\beta (\log_2 \lambda)$ in the time--frequency domain $(t, \log_2 \lambda)$ for different combinations of amplitude modulation rate $\alpha$ and frequency modulation scale $\beta$. Darker shades of red (resp.~blue) indicate higher positive (resp.~lower negative) values of the real part.}
\label{fig:spectrotemporal-wavelets}
\end{figure}

We define the fourth-order tensor
\begin{multline}
\mathbf{U}_2 \boldsymbol{x}(t, \lambda, \alpha, \beta) =
\big \vert \mathbf{U}_1 \boldsymbol{x} \overset{t}{\ast} \boldsymbol{\psi}_{\alpha} \overset{\log_2 \lambda}{\ast}  \boldsymbol{\psi}_{\beta} \big \vert (t, \lambda) \\
=
\Bigg\vert\iint\! \mathbf{U}_1 \boldsymbol{x} (\tau, s) \boldsymbol{\psi}_{\alpha}(t-\tau) \boldsymbol{\psi}_{\beta}(\log_2 \lambda - s) \,\mathrm{d}\tau \,\mathrm{d}s\Bigg\vert,
\label{eq:U2}
\end{multline}
which stacks convolutions in time and log-frequency with all wavelets $\boldsymbol{\psi}_\alpha (t)$ and $\boldsymbol{\psi}_\beta (\log_2 \lambda)$ followed by the complex modulus nonlinearity. 
Figure \ref{fig:spectrotemporal-wavelets} shows the interference pattern of the product $\boldsymbol{\psi}_{\alpha}(t - \tau) \boldsymbol{\psi}_\beta (\log_2 \lambda - s)$ for different combinations of time $t$, frequency $\lambda$, rate $\alpha$, and scale $\beta$.
We denote the multiindices $(\lambda, \alpha, \beta)$ resulting from such combinations as scattering \emph{paths} \cite{mallat2012cpam}.

\subsection{Averaging}

Because it is a convolutional operator in the time--frequency domain, the STRF is equivariant to temporal translation $t \mapsto t + \tau$ as well as frequency transposition $\lambda \mapsto 2^s \lambda$.
To produce musical metamers, we guarantee invariance to temporal translation up to some time lag $T$ and up to some frequency interval $F$.
To this aim, we define time--frequency scattering as the result of a local averaging of both $\mathbf{U_1} \boldsymbol{x} (t, \lambda)$ and $\mathbf{U}_2 \boldsymbol{x}(t, \lambda, \alpha, \beta)$ by a Gaussian low-pass filter $\boldsymbol{\phi}_T$ of cutoff frequency equal to $T^{-1}$, yielding
\begin{equation}
\mathbf{S}_1 \boldsymbol{x}(t, \lambda) =
\big( \mathbf{U}_1 \boldsymbol{x} \overset{t}{\ast} \boldsymbol{\phi}_T \overset{\log\lambda}{\ast} \boldsymbol{\phi}_F \big)(t, \lambda)
\label{eq:S1}
\end{equation}
and
\begin{equation}
\mathbf{S}_2 \boldsymbol{x}(t, \lambda, \alpha, \beta) =
\big(
\mathbf{U}_2 \boldsymbol{x} \overset{t}{\ast} \boldsymbol{\phi}_T \overset{\log_2 \lambda}{\ast} \boldsymbol{\phi}_F
\big)(t, \lambda, \alpha, \beta),
\label{eq:S2}
\end{equation}
respectively.

\subsection{Reconstruction}

By recursion over layers in the composition of wavelet modulus operators, the invertibility of wavelet modulus implies the invertibility of scattering transforms of infinite depth \cite{waldspurger2015phd}, up to a constant time shift of at most $T$.
However, the time--frequency scattering network presented here has a finite number of layers (i.e., two layers) and is therefore not exactly invertible.
Thanks to the theorem of exponential decay of scattering coefficients \cite{waldspurger2017exponential}, the residual energy that is present in deeper layers can be neglected on the condition that $T$ is small enough.

Starting from a colored Gaussian noise $\boldsymbol{y}_0 (t)$ whose power spectral density matches $\mathbf{S}_1 \boldsymbol{x} (t,\lambda)$, we refine it by additive updates of the form $\boldsymbol{y}_{n+1}(t) = \boldsymbol{y}_n (t) + \boldsymbol{u}_n (t)$, where the term $\boldsymbol{u}_n (t)$ is defined recursively as $\boldsymbol{u}_n (t) = m \times \boldsymbol{u}_n (t) + \mu_n \boldsymbol{\nabla}\mathbf{E}(\boldsymbol{y}_n)(t)$.
In subsequent experiments, the momentum term is fixed at $m=0.9$ while the learning rate is initialized at $\mu_0 = 0.1$ and modified at every step according to a ``bold driver'' heuristic \cite{sutskever2013icml}.

Like deep neural networks, scattering networks consist of the composition of linear operators (wavelet transforms) and pointwise nonlinearities (complex modulus).
Consequently, the gradient $\boldsymbol{\nabla}\mathbf{E}(\boldsymbol{y}_n)$ can be obtained by composing the Hermitian adjoints of these operator in the reverse order as in the direct scattering transform -- a method known as backpropagation.

We rely on the PyTorch backend of Kymatio to perform gradient backpropagation via reverse-mode automatic differentiation \cite{andreux2020kymatio}.
We outline the resulting automatic differentiation algorithm below.
First, Kymatio backpropagates the gradient of Euclidean loss for second-order scattering coefficients:
\begin{equation}
\boldsymbol{\nabla}\mathbf{U_2}\boldsymbol{y}(t,\lambda, \alpha, \beta) =
\Big((\mathbf{S_2}\boldsymbol{x}-\mathbf{S_2}\boldsymbol{y})
\overset{t}{\ast} \boldsymbol{\phi}\Big)
(t, \lambda, \alpha, \beta).
\end{equation}
Secondly, Kymatio backpropagates the second layer onto the first:
\begin{align}
\boldsymbol{\nabla}\mathbf{U_1}\boldsymbol{y}(t,\lambda) &=
    \Big((\mathbf{S_1}\boldsymbol{x}-\mathbf{S_1}\boldsymbol{y})
\overset{t}{\ast} \boldsymbol{\phi}\Big)
(t, \lambda, \alpha, \beta) \nonumber \\ &+
    \sum_{\alpha,\beta} \mathfrak{R}\Bigg( \Bigg[\dfrac{\big(\mathbf{U_1}\boldsymbol{y} \overset{t}{\ast} \boldsymbol{\bar{\psi}}_{\alpha} \overset{\log \lambda}{\ast} \boldsymbol{\bar{\psi}}_{\beta} \big)}{\big\vert\mathbf{U_1}\boldsymbol{y} \overset{t}{\ast} \boldsymbol{\bar{\psi}}_{\alpha} \overset{\log \lambda}{\ast} \boldsymbol{\bar{\psi}}_{\beta} \big\vert} \times \boldsymbol{\nabla}\mathbf{U_2}\boldsymbol{y} \Bigg]
    \overset{t}{\ast} \boldsymbol{\psi}_{\alpha} \overset{\log \lambda}{\ast} \boldsymbol{\psi}_{\beta} \Bigg)(t, \lambda, \alpha, \beta),
\end{align}
where the symbol $\mathfrak{R}(z)$ denotes the real part of the complex number $z$.
Lastly, Kymatio backpropagates the first layer into the waveform domain:
\begin{equation}
    \boldsymbol{\nabla}\mathbf{E}(\boldsymbol{y})(t) =
    \sum_{\lambda} \mathfrak{R}\Bigg( \Bigg[ \dfrac{\boldsymbol{y}\overset{t}{\ast}\boldsymbol{\psi}_{\lambda}}{\vert\boldsymbol{y}\overset{t}{\ast}\boldsymbol{\psi}_{\lambda}\vert} \times \boldsymbol{\nabla}\mathbf{U_1}\boldsymbol{y} \Bigg] \overset{t}{\ast} \boldsymbol{\psi}_{\lambda}\Bigg)(t)
\end{equation}
Observe that the signal above is null if $\mathbf{S_1}\boldsymbol{x}=\mathbf{S_1}\boldsymbol{y}$ and $\mathbf{S_2}\boldsymbol{x}=\mathbf{S_2}\boldsymbol{y}$; i.e., if $\boldsymbol{x}$ and $\boldsymbol{y}$ are musical metamers for the JTFS representation.

\section{Implementation}

\subsection{Filterbanks}
Let $(\boldsymbol{\psi}^{\mathrm{tm},1}_{0}, \ldots, \boldsymbol{\psi}^{\mathrm{tm},1}_{N_1-1})$ be a finite sequence of complex-valued band-pass filters with center frequencies $1>\xi^{\mathrm{tm},1}_{0}>\ldots>\xi^{\mathrm{tm},1}_{N_1-1}>0$ and bandwidths $\frac{1}{2}>\sigma^{\mathrm{tm},1}_{0}>\ldots>\sigma^{\mathrm{tm},1}_{N_1-1}>0$, where the sample rate is set equal to $1$ by convention.
We refer to \cite{lostanlen2018relevance} for the rationale behind the values of $\xi_0^{\mathrm{tm},1}$ and $\sigma_0^{\mathrm{tm},1}$.

Given a scale $J$ and a dimensionless hyperparameter $Q^{\mathrm{tm},1}$, center frequencies and bandwiths decrease according to a geometric progression, while maintaining a quality factor of $Q^{\mathrm{tm},1}$; until reaching an ``elbow'', defined as a lower bound on the bandwidth.
After the elbow, the progression of center frequencies is arithmetic, while bandwidths remain at the lower bound.
Thus, filters have a constant-bandwidth property in the lower-frequency range and a constant-$Q$ property in the upper-frequency range.
This construction was first proposed in \cite{anden2011multiscale} and is reminiscent of the construction of the mel scale in HTK\footnote{https://librosa.org}.
We reproduce an excerpt of the Kymatio code below.

\begin{lstlisting}
sigma0 = 0.1
r_psi = math.sqrt(0.5)

xi = max(1. / (1. + math.pow(2., 3. / Q)), 0.35)
factor = 1. / math.pow(2, 1. / Q)
sigma = (1 - factor) / (1 + factor) * xi / math.sqrt(2 * math.log(1. / r))
sigma_min = sigma0 / 2**J

if sigma <= sigma_min:
    xi = sigma
else:
    yield xi, sigma
    # High-frequency (constant-Q) region
    while sigma > (sigma_min * math.pow(2, 1 / Q)):
        xi /= math.pow(2, 1 / Q)
        sigma /= math.pow(2, 1 / Q)
        yield xi, sigma

# Low-frequency (constant-bandwidth) region
elbow_xi = xi
for q in range(Q-1):
    xi -= 1/Q * elbow_xi
    yield xi, sigma_min
\end{lstlisting}
Another filterbank is built for the second layer: $(\boldsymbol{\psi}^{\mathrm{tm},2}_{0}, \ldots, \boldsymbol{\psi}^{\mathrm{tm},2}_{N_2-1})$, with its own hyperparameters $Q^{\mathrm{tm},2}$.
Lastly, we build a filterbank for frequential scattering: $(\boldsymbol{\psi}^{\mathrm{fr},1}_{0}, \ldots, \boldsymbol{\psi}^{\mathrm{fr},1}_{N_{\mathrm{fr}}-1})$, with its own hyperparameters $Q^{\mathrm{fr},1}$ and $J^{\mathrm{fr},1}$.

\subsection{First layer}
Given an audio signal $\boldsymbol{x}$, we extract its time--frequency representation by convolution with each filters $\boldsymbol{\psi}^{\mathrm{tm},1}_{n_1}$ where $0 \leq n_1 < N_1$.
For reasons of computational efficiency, we map the signal to the Fourier domain, via a real-input direct fast Fourier transform (RFFT).
The filtering itself is an element-wise product between complex-valued vectors, implemented with the CDGMM routine.
Furthermore, the filtering is immediately followed by a subsampling operation, implemented in the Fourier domain by reshaping and summation.
The subsampling factor $2^{k_1}$, depends on the resolution of the cutoff frequency of the filter $\boldsymbol{\psi}^{\mathrm{tm},1}_{n_1}$, i.e., $j_1$; and on the choice of stride, i.e., $T$.
We reproduce an excerpt of the Kymatio code below.
\begin{lstlisting}
U_0_hat = backend.rfft(U_0)
for n1 in psi1:
    j1 = psi1[n1]['j']
    k1 = min(j1, log2_stride)
    U_1_c = backend.cdgmm(U_0_hat, psi1[n1]['levels'][0])
    U_1_hat = backend.subsample_fourier(U_1_c, 2 ** k1)
    U_1_c = backend.ifft(U_1_hat)
    U_1_m = backend.modulus(U_1_c)
    U_1_hat = backend.rfft(U_1_m)
    U_1_hats.append({'coef': U_1_hat, 'j': (j1,), 'n': (n1,)})
\end{lstlisting}
To reduce the influence of boundary effects, this operation is preceded by padding on both sides, and followed by unpadding on both sides, not shown in the code above.

\subsection{Second layer, time variable}
Contrary to Kymatio's time scattering module, which relies on a depth-first search over scattering paths for memory efficiency, time scattering in Kymatio's JTFS module is done according to a width-first search.
In other words, the outer loop corresponds to the deeper path index ($n_2$) whereas the inner loop corresponds to the shallower path index ($n_1$).
This is because all the first-order paths corresponding to a given second-order paths must be available in memory when performing scattering over the log-frequency variable (see next subsection).
We reproduce an excerpt of the Kymatio code below.
\begin{lstlisting}
psi2 = filters[2]
for n2 in range(len(psi2)):
    j2 = psi2[n2]['j']
    Y_2_list = []

    for U_1_hat in U_1_hats:
        j1 = U_1_hat['j'][0]
        k1 = min(j1, log2_stride) if average_local else j1

        if j2 > j1:
            k2 = min(j2-k1, log2_stride) if average_local else (j2-k1)
            U_2_c = backend.cdgmm(U_1_hat['coef'], psi2[n2]['levels'][k1])
            U_2_hat = backend.subsample_fourier(U_2_c, 2 ** k2)
            U_2_c = backend.ifft(U_2_hat)
            Y_2_list.append(U_2_c)

    if len(Y_2_list) > 0:
        # Stack over the penultimate dimension with shared n2.
        # Y_2 is a complex-valued 3D array indexed by (batch, n1, time)
        Y_2 = backend.stack(Y_2_list, 2)

        # Y_2 is a stack of multiple n1 paths so we put (-1) as placeholder.
        # n1 ranges between 0 (included) and n1_max (excluded), which we store
        # separately for the sake of meta() and padding/unpadding downstream.
        yield {'coef': Y_2, 'j': (-1, j2), 'n': (-1, n2), 'n1_max': len(Y_2_list)} 
\end{lstlisting}

\subsection{Second layer, frequency variable}
Since band-pass filters $\boldsymbol{\psi}^{\mathrm{tm},1}_{0}, \ldots, \boldsymbol{\psi}^{\mathrm{tm},1}_{N_1-1}$ are complex-valued, scattered signals $(\vert\boldsymbol{x}\ast\boldsymbol{\psi}^{\mathrm{tm},1}\vert\ast\boldsymbol{\psi}^{\mathrm{tm},2})$ are also complex-valued.
Thus, it is necessary to include negative frequencies as well as positive frequencies.
For this purpose, we flip the sign of all center frequencies $\xi_{n}^{\mathrm{fr},1}$ while maintaining the same bandwidths $\sigma_{n}^{\mathrm{fr},1}$; an operation known as wavelet \emph{spinning}.
We reproduce an excerpt of the Kymatio code below.
\begin{lstlisting}
def spin(filterbank_fn, filterbank_kwargs):
    def spinned_fn(J, Q, **kwargs):
        yield from filterbank_fn(J, Q, **kwargs)
        for xi, sigma in filterbank_fn(J, Q, **kwargs):
            yield -xi, sigma
    return spinned_fn, filterbank_kwargs
\end{lstlisting}
Importantly, spinning is only necessary if $\alpha\neq0$: indeed, the case $\alpha=0$ corresponds to a temporal low-pass filtering of the scalogram, which produces a real-valued input. 
We accommodate both subcases in Kymatio via the \texttt{spinned} Boolean argument:
\begin{lstlisting}
def frequency_scattering(X, backend, filters_fr, log2_stride_fr,
        average_local_fr, spinned):

    # Unpack filters_fr list
    phi, psis = filters_fr
    log2_F = phi['j']

    # Swap time and frequency axis
    X_T = backend.swap_time_frequency(X['coef'])

    # Zero-pad frequency domain
    pad_right = phi['N'] - X['n1_max']
    X_pad = backend.pad_frequency(X_T, pad_right)

    # Spinned case switch
    if spinned:
        # Complex-input FFT
        X_hat = backend.cfft(X_pad)
    else:
        # Real-input FFT
        X_hat = backend.rfft(X_pad)
        # Restrict to nonnegative spins
        psis = filter(lambda psi: psi['xi']>=0, psis)

    for n_fr, psi in enumerate(psis):
        j_fr = psi['j']
        spin = np.sign(psi['xi'])
        k_fr = min(j_fr, log2_stride_fr) if average_local_fr else j_fr
        Y_fr_hat = backend.cdgmm(X_hat, psi['levels'][0])
        Y_fr_sub = backend.subsample_fourier(Y_fr_hat, 2 ** k_fr)
        Y_fr = backend.ifft(Y_fr_sub)
        Y_fr = backend.swap_time_frequency(Y_fr)
        # If not spinned, X['n']=S1['n']=(-1,) is a 1-tuple.
        # If spinned, X['n']=Y2['n']=(-1,n2) is a 2-tuple.
        # In either case, we elide the "-1" placeholder and define the new 'n'
        # as (X['n'][1:] + (n_fr,)), i.e., n=(n_fr,) is not spinned
        # and n=(n2, n_fr) if spinned. This 'n' tuple is unique.
        yield {**X, 'coef': Y_fr, 'n': (X['n'][1:] + (n_fr,)),
            'j_fr': (j_fr,), 'n_fr': (n_fr,), 'n1_stride': (2**j_fr),
            'spin': spin}
\end{lstlisting}

\subsection{Main algorithm}
The main JTFS algorithm has only six lines of code, thanks to the \texttt{yield} coroutine and the \texttt{yield from} expression, added in Python 3.3.
\begin{lstlisting}
def joint_timefrequency_scattering(U_0, backend, filters, log2_stride,
         average_local, filters_fr, log2_stride_fr, average_local_fr):
    """
    Yields
    ------
    # Zeroth order
    if average_local:
     * S_0 indexed by (batch, time[log2_T])
    else:
     * U_0 indexed by (batch, time)

    # First order
    for n_fr < len(filters_fr):
     * Y_1_fr indexed by (batch, n1[n_fr], time[log2_T]), complex-valued,
         where n1 has been zero-padded to size N_fr before convolution

    # Second order
    for n2 < len(psi2):
     for n_fr < len(filters_fr):
         * Y_2_fr indexed by (batch, n1[n_fr], time[n2]), complex-valued,
             where n1 has been zero-padded to size N_fr before convolution
    """
    time_gen = time_scattering_widthfirst(
        U_0, backend, filters, log2_stride, average_local)
    yield next(time_gen)
    
    # First order: S1(n1, t) = (|x*psi_{n1}|*phi)(t[log2_T])
    S_1 = next(time_gen)
    
    # Y_1_fr_{n_fr}(n1, t[log2_T]) = (|x*psi_{n1}|*phi*psi_{n_fr})(t[log2_T])
    yield from frequency_scattering(S_1, backend,
        filters_fr, log2_stride_fr, average_local_fr, spinned=False)
    
    # Second order: Y_2_{n2}(t[log2_T], n1[j_fr])
    #                   = (|x*psi_{n1}|*psi_{n2})(t[j2], n1[j_fr])
    for Y_2 in time_gen:
    
        # Y_2_fr_{n2,n_fr}(t[j2], n1[j_fr])
        #     = (|x*psi_{n1}|*psi_{n2}*psi_{n_fr})(t[j2], n1[j_fr])
        yield from frequency_scattering(Y_2, backend, filters_fr,
            log2_stride_fr, average_local_fr, spinned=True)
\end{lstlisting}

\section{Related work}
\subsection{Applications to audio classification}
The first implementation of JTFS was made in 2014 by Joakim Andén and was written in MATLAB\footnote{\url{https://github.com/scatnet/scatnet}}.
In 2015, Vincent Lostanlen wrote another MATLAB implementation enabling custom routines for gradient backpropagation \footnote{\url{https://github.com/lostanlen/scattering.m}}.
The first JTFS article relied on the first implementation for speech classification and on the second implementation for texture synthesis of bird sounds \cite{anden2015joint}.
A mathematical link between JTFS and deep convolutional network was outlined by \cite{mallat2016understanding} and formally built by \cite{anden2019joint}, with new applications to the supervised classification of music instruments and urban sounds.
Since then, JTFS has been applied to other domains: playing techniques from Chinese bamboo flute \cite{wang2020playing}, bioacoustics \cite{wang2022joint}, and medical acoustics \cite{sharma2023time}.
Another publication has shown that Euclidean distances in JTFS feature space predict auditory judgments of timbre similarity within a large vocabulary of instrumental playing techniques, as collected from a group of 31 professional composers and non-expert music listeners \cite{lostanlen2021time}.
An article for the French academy of sciences, written on the anniversary of Joseph Fourier's birth, discusses the historical background of JTFS \cite{lostanlen2019fourier}.

\subsection{Use in computer music}
The earliest use of JTFS in computer music was \emph{FAVN}, an opera which was commissioned to composer Florian Hecker by the Alte Oper in Frankfurt in 2016, followed by an exhibition at the Kunsthalle in Vienna in 2017.
The catalog of this exhibition discusses the underpinnings of JTFS in terms of programming language theory, as well as connections with auditory neurophysiology \cite{lostanlen2018time}.
This discussion was expanded in \cite{lostanlen2019shape} on the occasion of \emph{The Shape of RemiXXXes to Come}: i.e., a Lorenzo Senni record, released by Warp Records, containing a JTFS-based remix by Florian Hecker.
JTFS was also used in other pieces, such as \emph{Synopsis Seriation} \cite{lostanlen2021synopsis}, released by Urbanomic, and \emph{Resynthese FAVN}, released as a 10-CD box set by Blank Forms.
The use of JTFS in music has been discussed by art historian Ina Blom \cite{blom2021sound} and by philosopher Amy Ireland \cite{ireland2024lado}.

\subsection{Kymatio implementation}
JTFS was re-implemented in the Python language by Mathieu Andreux in 2018 \cite{andreux2018music}, with an application to generative audio models.
The next year, the Kymatio consortium was officially created, with the goal of providing a stable, portable, and efficient implementation of scattering transforms in Python\footnote{\url{https://kymat.io}}.
The development of time scattering in Kymatio was led by Andén, Andreux, and Lostanlen, with important contributions by John Muradeli, Changhong Wang, and Philip Warrick \cite{andreux2020kymatio}.
A first implementation of GPU-enabled JTFS from Kymatio's Pytorch routines was written by John Muradeli, under the name of wavespin\footnote{\url{https://github.com/OverLordGoldDragon/wavespin/tree/dafx2022-jtfs}}, with new applications to manifold learning and music instrument classification \cite{muradeli2022differentiable}.
The current implementation of JTFS in Kymatio was written by Vincent Lostanlen, Cyrus Asfa, Han Han, Changhong Wang, Muawiz Chaudhary, and Joakim Andén\footnote{\url{https://github.com/kymatio/kymatio}}.
Since then, Chris Mitcheltree has reimplemented JTFS in PyTorch for pedagogical purposes\footnote{\url{https://github.com/christhetree/jtfst_implementation}}.

Reverse-mode automatic differentiation in Kymatio has been successfully applied to perceptual sound matching \cite{vahidi2023mesostructures}.
The combination of JTFS, differentiable physical modeling synthesis, and Riemannian geometry has led to an original technique for training neural networks to solve inverse problems, known as perceptual--neural--physical sound matching \cite{han2023perceptual,han2024learning}.
This is a kind of model-based deep learning for music information research---we refer to \cite{richard2024model} for an introduction.

\subsection{Spectrotemporal receptive fields}
Besides JTFS, there are other algorithms which extract spectrotemporal modulations at various scales and rates.
As early as 1998, \cite{chi1999spectro} have implemented spectro-temporal modulation transfer functions so as to fit neurophysiological responses to ripple noise stimuli, as previously measured in the auditory cortex of ferrets \cite{kowalski1996analysis}.
We refer to \cite{chi2005multiresolution} for a complete mathematical description of this implementation, known as ``full cortical model'' \cite{patil2012music}, and to \cite{lindeberg2015plos} for a review of applications.
A MATLAB implementation of STRF, including the synthesis of sounds from time-averaged (but not frequency-averaged) coefficients, was achieved by Sam Norman-Haigneré \cite{norman2018neural} \footnote{\url{https://github.com/snormanhaignere/spectrotemporal-synthesis-v2}}.
We refer to \cite{siedenburg2016comparison} for an introduction to STRF in the interdisciplinary context of music cognition and music information retrieval (MIR).
Recently, STRF has been applied to other tasks, such as bioacoustics \cite{min2024few} and industrial acoustics \cite{li2024machine}.

Another implementation of STRF was proposed in 2012 by \cite{schadler2012spectro}, under the name of Gabor filterbanks (GBFB), with applications to automatic keyword spotting in noisy environments.
In music cognition, STRF is sometimes known under the name of ``modulation power spectrum'', with applications to timbre perception research \cite{thoret2017perceptually}.

\section{Conclusion}
We have described a method to generate musical metamers from any audio recording.
Our method is based on joint time--frequency scattering in Kymatio, an open-source software in Python which enables GPU computing and automatic differentiation.
The advantage of our method is that it does not require any manual preprocessing, such as transcription, beat tracking, or source separation.
Hence, it is suitable to any kind of audio signal, without need to postulate the presence of ``Western classical'' cues such as tonality, counterpoint, meter, or theme.
Lastly, because the resynthesis operates by iterative refinement from a random initial guess, it is compatible with the methodological guidelines of reproducible research: by running the algorithm anew, one may generate an infinite number of \emph{independent} metamers without compromising the practitioner's ability to conduct counterfactual reasoning.
In future work, we plan to use the generated metamers in the context of music cognition to as to test Honing's hypothesis of multicomponent contour perception at various spectrotemporal scales.

\def\refname{R\MakeLowercase{eferences}}

\bibliographystyle{IEEEtran}
\bibliography{lostanlen2024metamerism}

\end{document}